\pdfoutput=1

\documentclass[10pt, conference, letterpaper]{IEEEtran}

\usepackage{color, colortbl}
\definecolor{Gray}{gray}{0.9}
\definecolor{LightCyan}{rgb}{0.88,1,1}
\definecolor{LightRed}{rgb}{1,0.88,0.88}
\definecolor{LightBlue}{rgb}{0.12,0.56,0.8}
\usepackage[table]{xcolor}

\usepackage[first=0,last=9]{lcg}

\usepackage{xfrac} 

\usepackage{cite}

\usepackage{graphicx}
\usepackage{amsmath}
\usepackage[linesnumbered,ruled,vlined]{algorithm2e}
\SetKwProg{Fn}{Function}{}{}

\SetCommentSty{mycommfont}

\makeatletter
\newcommand{\removelatexerror}{\let\@latex@error\@gobble}
\makeatother

\usepackage{array}

\ifCLASSOPTIONcompsoc
  \usepackage[caption=false,font=normalsize,labelfont=sf,textfont=sf]{subfig}
\else
  \usepackage[caption=false,font=footnotesize]{subfig}
\fi
\usepackage{dblfloatfix}    

\usepackage{url}

\IEEEoverridecommandlockouts
\IEEEpubid{%
   \makebox[\columnwidth]{ISBN 978-3-903176-08-9 © 2018 IFIP\hfill}%
   \hspace{\columnsep}%
   \makebox[\columnwidth]{ }%
}
\begin{document}
\bstctlcite{IEEEexample:BSTcontrol}

%
\title{Cellular Controlled Delay TCP (C2TCP)}


\author{\IEEEauthorblockN{Soheil Abbasloo, Tong Li, Yang Xu, H. Jonathan Chao}
\IEEEauthorblockA{New York University\\Email: \{ab.soheil, tl1914, yang, chao\}@nyu.edu}}


%


\maketitle

\begin{abstract}
Cellular networks have special characteristics including highly variable channels, fast fluctuating capacities, deep per user buffers, self-inflicted queuing delays, radio uplink/downlink scheduling delays, etc. These distinguishing properties make the problem of achieving low latency and high throughput in cellular networks more challenging than in wired networks. That's why in this environment, TCP and its flavors, which are generally designed for wired networks, perform poorly. 

To cope with these challenges, we present C2TCP, a flexible end-to-end solution targeting interactive applications requiring high throughput and low delay in cellular networks. C2TCP stands on top of loss-based TCP and brings it delay sensitivity without requiring any network state profiling, channel prediction, or complicated rate adjustment mechanisms. The key idea behind C2TCP is to absorb dynamics of unpredictable cellular channels by investigating local minimum delay of packets in a moving time window and react to the cellular network's capacity changes very fast. 

Through extensive trace-based evaluations using traces from five commercial LTE and 3G networks, we have compared performance of C2TCP with various TCP variants, and state-of-the-art schemes including BBR, Verus, and Sprout. Results show that on average, C2TCP outperforms these schemes and achieves lower average and 95th percentile delay for packets.
\end{abstract}


%
\IEEEpeerreviewmaketitle

\section{Introduction}
Cumulative data traffic growth in cellular networks has increased more than 1200\% over recent five-year period and in the first quarter of 2017, total cellular internet traffic reached to nearly 9500 PetaByte per month globally~\cite{mobile-stat}. This growing mode of internet access on the one hand has provided opportunities for cellular network carriers with more demand for new applications like augmented reality, virtual reality, online gaming, real time video streaming, and real time remote health monitoring. On the other hand it has brought new challenges for cellular carriers due to ultra low latency and high throughput requirements of those interactive applications. 

Cellular networks differ noticeably from their wired counter parts. They experience highly variable channels, fast fluctuating capacities, self-inflicted queuing delays, stochastic packet losses, and radio uplink/downlink scheduling delays. These differences make the problem of achieving low latency and high throughput in cellular networks more challenging than in wired networks. TCP and its variants which are the main congestion control mechanisms to control the delay and throughput of flows are known to perform poorly in cellular networks~\cite{sprout, ex-tcp, verus, lte_depth, bufferbloat2}.

In this paper we present \textit{C2TCP}, a \textit{\textbf{C}ellular \textbf{C}ontrolled delay \textbf{TCP}} to address mentioned challenges in cellular networks for achieving low delay and high throughput. Our main philosophy is that achieving good performance does not necessarily comes from complex rate calculation algorithms or complicated channel modelings.\footnote{It is already a known fact that predicting cellular channels is hard ~\cite{verus,3g-channel}} The key idea behind C2TCP's design is to absorb dynamics of unpredictable cellular channels by investigating local minimum of packets' delay in a moving time window. Doing that, C2TCP stands on top of a loss-based TCP such as Cubic~\cite{cubic} and brings it delay sensitivity. There is no network state profiling, channel prediction, or any complicated rate adjustments mechanisms involved. 

There is always a trade-off between achieving lowest delay and getting highest throughput. J. Jaffe in \cite{jaf} has proved that no distributed algorithm can converge to the operation point in which both the minimum RTT and maximum throughput are achieved. Considering that trade-off, C2TCP provides a flexible end-to-end solution which allows applications to choose their level of delay sensitiveness, even after the start of their connection. 

We have implemented C2TCP in Linux Kernel 4.13, on top of Cubic, conducted extensive trace-driven evaluations (detailed in section~\ref{eval}) using data collected in prior work (\cite{mahi} and \cite{sprout}) from 5 different commercial cellular networks in Boston (T-Mobile's LTE and 3G UMTS, AT\&T's LTE, and Verizon's LTE and 3G 1xEV-DO)  in both directions, and compared performance of C2TCP with several TCP variants (including Cubic~\cite{cubic}, NewReno~\cite{newreno}, and Vegas~\cite{vegas}) and different state-of-the-art end-to-end schemes including BBR~\cite{bbr}, Sprout~\cite{sprout}, and Verus~\cite{verus}. Our results show that C2TCP outperforms these end-to-end schemes and achieves lower average and 95th percentile delays for packets. In particular, on average, Sprout, Verus, and BBR have $3.41\times$, $10.36\times$, and $1.87\times$ higher average delays and $1.44\times$, $27.36\times$, and $2.06\times$ higher 95th percentile delays compared to C2TCP, respectively. This great delay performance comes at little cost in throughput. For instance compared to Verus (which achieves the highest throughput among those 3 state-of-the-art schemes), C2TCP's throughput is only $0.22\times$ less.

Also, in section~\ref{eval-codel}, we compared our end-to-end solution, C2TCP, with CoDel~\cite{codel}, an AQM scheme that requires modification on carriers network, and show that C2TCP can outperform CoDel too. Moreover, we examined fairness of C2TCP, compared it with several other algorithms, and showed that it provides good fairness properties. Finally, we investigated the loss resiliency of C2TCP in case of stochastic packet losses unrelated to congestion in cellular networks. Among algorithms that we examined only C2TCP, BBR, and Vegas show good resiliency in high rates of packet losses.  

\section{Motivations and Design Decisions}
\textbf{Flexible End-to-End Approach:} One of the key distinguishing features of cellular networks is that cellualr carriers generally provision deep per user queues in both uplink and downlink directions at base station (BS). This leads to issues such as self-inflicted queuing delay~\cite{sprout} and bufferbloat~\cite{bufferbloat,bufferbloat2}. A traditional solution for these issues is using AQM schemes like RED~\cite{red}; however, correct parameter tuning of these algorithms to meet requirements of different applications is challenging and difficult. Although newer AQM algorithms such as CoDel~\cite{codel} can solve the tuning issue, they design queues from scratch, so deploying them in network comes with huge CAPEX cost. In addition, in-network schemes lack flexibility. They are based on ``one-setting-for-all-applications'' concept and won't consider that different type of applications might have different delay and throughput requirements. Moreover, with new trends and architectures such as mobile content delivery network (MCDN) and mobile edge computing (MEC)~\cite{mec}, content is being pushed close to the end-users. So, from the latency point of view, problem of potential large control feedback delay of end-to-end solutions diminishes if not disappears. Motivated by these shortcomings and new trends, we seek a ``\textit{flexible end-to-end}'' solution without tuning difficulties.

\textbf{Simplicity:} Cellular channels often experience fast fluctuations and widely variable capacity changes over both short and long timescales~\cite{verus}. This property along with several complex lower layer state machine transitions\cite{lte_depth}, complicated interactions between user equipment (UE) and BS \cite{lte-book}, and scheduling algorithms used in BS to allocate resources for users through time which are generally unknown for end-users make cellular channels hard to be predictable if not \textit{unpredictable}~\cite{verus,3g-channel}. These complexities and unpredictability nature of channels motivates us to avoid using any channel modeling/prediction and to prevent adding more complicity to cellular networks. We believe that performance doesn't always come from complexity. Therefore, we seek ``\textit{simple yet powerful}'' approaches to tackle congestion issue in cellular networks.

\textbf{Network as a Black-Box:} In cellular networks, source of delay is vague. End-to-end delay could be due to either self-inflicted queuing delays in BS, delays caused by BS' scheduling decisions in both directions, or downlink/uplink channel fluctuations. Although providing feedback from network to users can guide them to detect the main source of delay, any new design based on requiring new feedback from network needs modifications on cellular networks. However, this comes at the CAPEX cost for cellular carriers. Therefore, we will look at cellular network as a ``\textit{black-box}'' which doesn't provide us any information about itself directly.

\section{C2TCP's Algorithm}   
\subsection{The Good and The Bad Conditions}
Inspired by CoDel~\cite{codel} and Vegas~\cite{vegas} designs, we define that network is in ``\textit{good-condition}'' when minimum RTT observed in a monitoring time duration (called \textit{Interval}) remains below a \textit{Target} delay. If RTT of a packet goes above the Target and if RTTs of next packets never fall below the Target in the next \textit{Interval}, we say network is in a \textit{bad-condition}. So, having any RTT less than \textit{Target} shows a good-condition at least for the next \textit{Interval}, while instantaneous RTTs bigger than \textit{Target} doesn't necessarily show a ``\textit{bad-condition}''. 
Intuitively, this definition comes from the fact that one of the normal responsibilities of queues in the network is to absorb burst of traffic, so it is normal to have some temporary increase in RTT of packets. However, when more packets experience large RTTs, most likely there is something wrong in the network. Hence, history of delay should be considered as important as the current delay. 

\begin{figure}[!t]
\centering
\includegraphics[width=0.48\textwidth,height=1.2in]{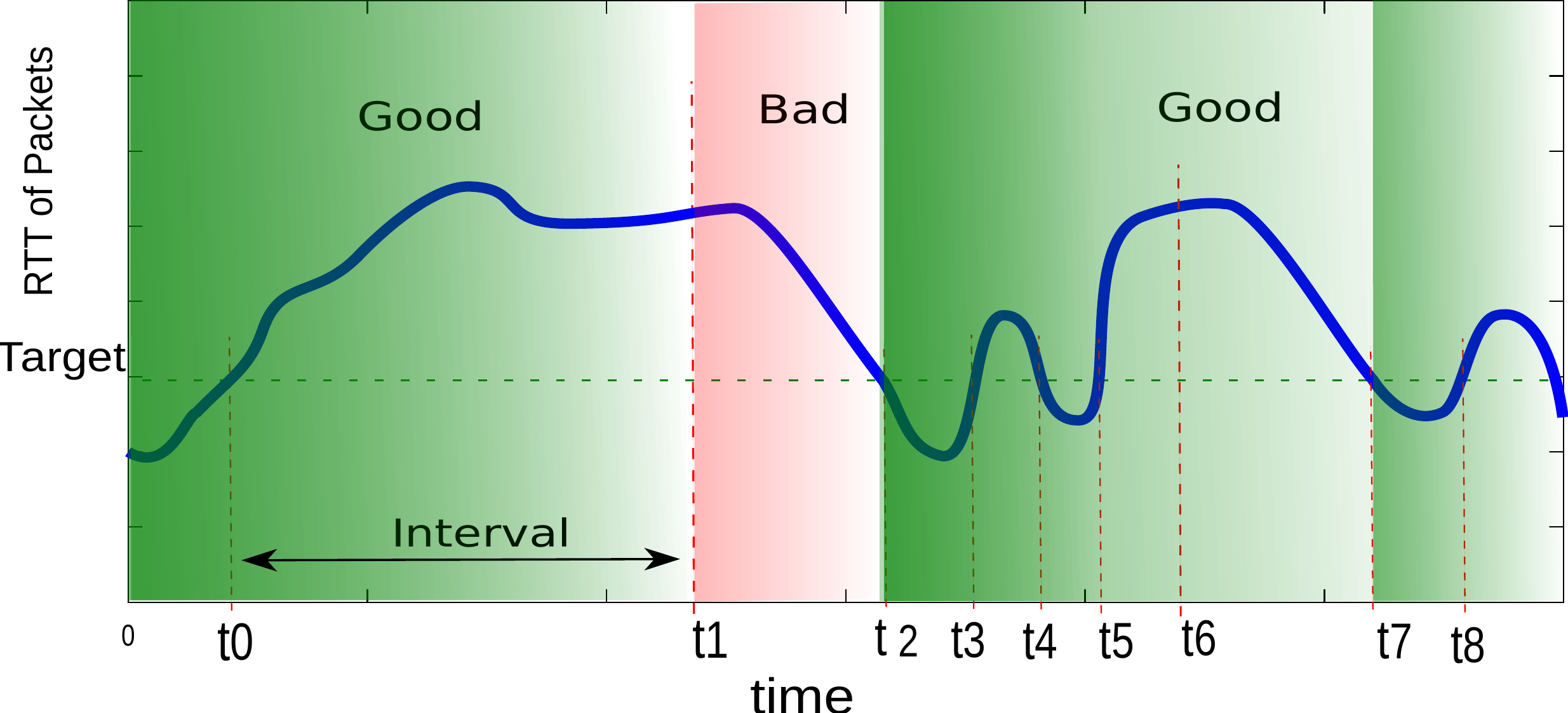}
\caption{The Good and the bad conditions}\label{fig_good-bad}
\end{figure}

For instance, consider Fig.~\ref{fig_good-bad} which shows sample RTTs of packets through time. At $t_0$ RTT of a packet goes beyond the Target value and till $t_1=t_0+Interval$ no packet experiences RTT less than Target. So, at $t_1$ a bad-condition is detected. This bad-condition continues till $t_2$ when RTT goes below Target value indicating detection of a good-condition. 
At $t_3$ RTT goes above Target but since RTT becomes less than Target at $t_4$ ($< Interval+t_3$), we still remain in good-condition. Likewise we remain in good-condition for the next time slots.  

Note that the delay responses of packets in $[t_1,t_2]$ and $[t_6,t_7]$ periods are identical. However, since the history of delay is different at $t_1$ and $t_6$, those two periods have been identified differently (first one is in a bad-condition, while the second one is in a good-condition). This example shows how we can use our simple definition to qualitatively get a sense of history without recording history of delay. 

\subsection{Main Logic}
As Algorithm~\ref{algorithm} shows, C2TCP's main logic will be triggered each time a new acknowledgment is received at the source. After detecting a bad-condition, the main question is that \textit{if we had an in-network AQM algorithm able to detect the bad-condition, what would have it done to inform a loss-based TCP?} The answer is that it would have simply dropped the packet and caused TCP to do a harsh back-off by setting congestion window to one. So, the key idea of C2TCP is to emulate such an impact without requiring that imaginary AQM scheme in the network. When bad-condition is detected at source, C2TCP overwrites the decision of TCP and forces congestion window to be reset to one. Each time a bad-condition is detected, the next monitoring time interval is decreased in proportion to $\frac{1}{\sqrt{n}}$ where $n$ is number of consecutive back-offs since detecting the current bad-condition using the well-known relationship of drop rate to throughput to get a linear change in throughput~\cite{sqrt,sqrt2}. 

On the other hand, RTTs smaller than Target show room for applications to increase their throughput at cost of increase in their delay. Therefore, we break the good-condition into two phases:  1- When RTTs are smaller than Target and 2- When RTTs are larger than Target (waiting phase). In waiting phase (lines 9-12 in Algorithm~\ref{algorithm}) , C2TCP doesn't change the congestion window calculated by the loss-based TCP, which is used as base for C2TCP, and waits for transition to either bad-condition or another first phase of the good-condition. However, in the first phase, C2TCP increases the congestion window additively using equation~\ref{eq_cwnd} (lines 5-9 in Algorithm~\ref{algorithm}). This increase is in addition to the increase that the loss-based TCP normally does. The choice of this additive increase is to follow the well-known AIMD (Additive Increase Multiplicative Decrease) property which ensures that C2TCP's algorithm still achieves fairness among connections~\cite{aimd}. We have examined C2TCP's fairness in more detail in section~\ref{sec_fair}. 

\begin{equation} 
\label{eq_cwnd} 
Cwnd_{new} = Cwnd_{current}+\frac{Target}{rtt_{current}}
\end{equation}

\begin{figure}[!t]
 \removelatexerror
  \begin{algorithm}[H]
  \DontPrintSemicolon
   \caption{C2TCP's Main Algorithm at Sender}
   \label{algorithm}
	\Fn(\tcp*[h]{process a new received Ack}){pkts\_acked()}{ 
	...	\tcc*[l]{default loss-based TCP code block} 	
		$rtt \longleftarrow current\_rtt$\;
		$now \longleftarrow current\_time$\;
		\uIf(){$rtt < Target$}{ \tcc*[l]{good-condition}
			$Cwnd \mathrel{+}= \frac{Target}{rtt}$\;
			$first\_time \longleftarrow true$\;
			$num\_backoffs \longleftarrow 1$
		}
		\uElseIf{first\_time}{\tcc*[l]{waiting phase}
			$next\_time \longleftarrow now+Interval$\;
			$first\_time \longleftarrow false$\;
		}	
		\uElseIf(){$now > next\_time$}{\tcc*[l]{bad-condition}
			$next\_time \longleftarrow now+\sfrac{Interval}{\sqrt{num\_backoffs}}$\;
			$num\_backoffs \mathrel{+}+$\;
			\tcc*[l]{setting ssthresh using default TCP function which normally recalculates it in congestion avoidance phase}
			$ssthresh \longleftarrow recalc\_ssthresh()$\;
			$Cwnd \longleftarrow 1$\;
		}
	 }
  \end{algorithm}
\end{figure}
\subsection{Why It Works}
To show the improvements achieved by C2TCP and discuss the reasons, we compare the performance of C2TCP implemented on top of Cubic with Cubic following instructions described in section~\ref{eval}. Fig.~\ref{fig_why} shows 60 seconds of varying capacity of a cellular link (Verizon LTE network in downlink direction measured in Boston by prior work~\cite{sprout}) and delay/throughput performance of C2TCP and Cubic.

\subsubsection{Avoiding excessive packet sending} 
\begin{figure}[!t]
\centering
\includegraphics[width=0.5\textwidth,height=1.7in]{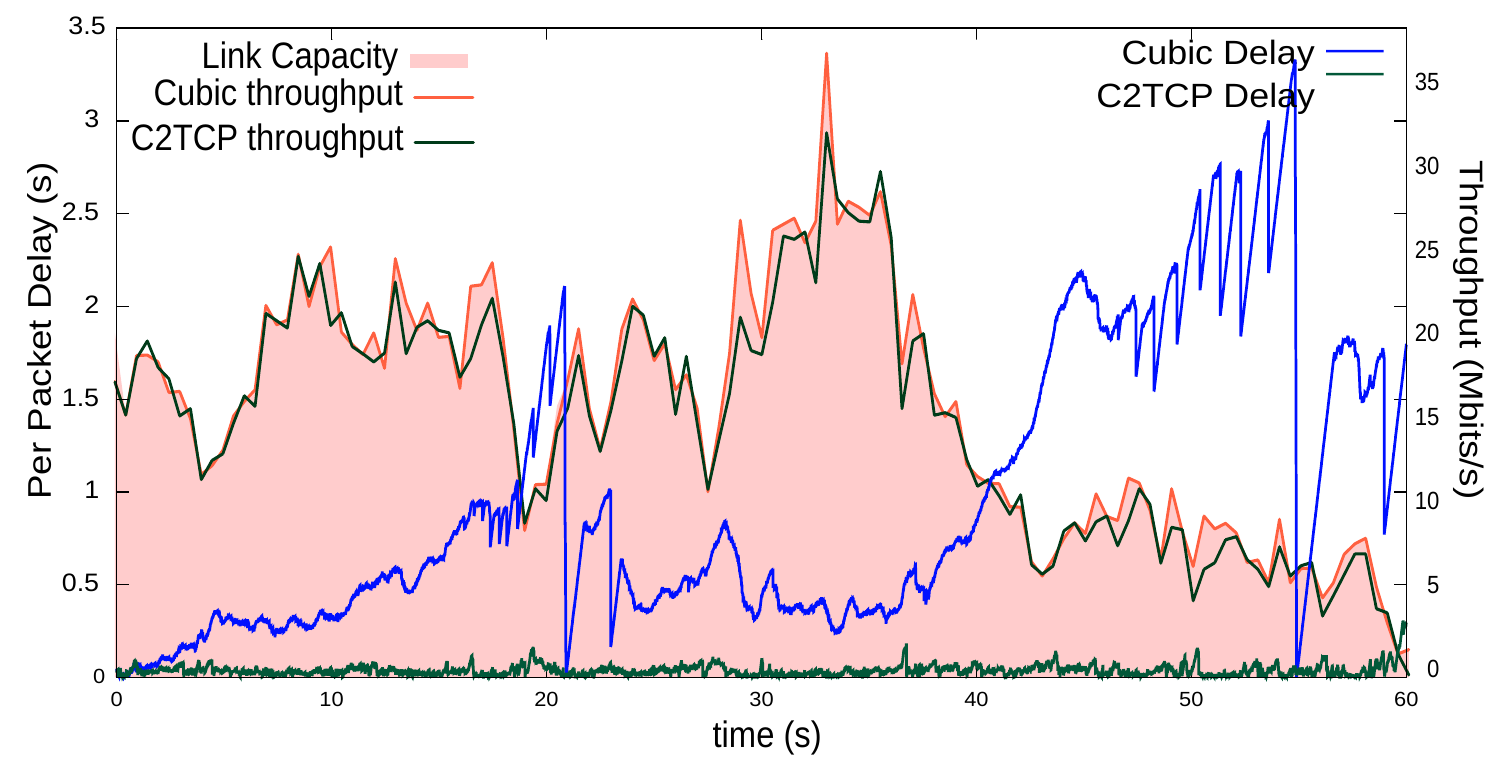}
\caption{Delay Response of Cubic and C2TCP}
\label{fig_why}
\end{figure}

Due to variations in link capacity and deep per user buffers, Cubic's delay performance is poor, specially when there is a sudden drop in capacity of link after experiencing good capacity (for instance, look at $[15s-20s]$ and $[35s-45s]$ time periods in Fig.~\ref{fig_why}). However, C2TCP always perform very well regardless of the fast link fluctuations. The key reason is that, C2TCP always keeps \textit{proper} amount of packets in the queues so that on the one hand, it avoids queue buildup and increase in the packet delay and on the other hand, it achieves high utilization of the cellular access link when either channel quality becomes good or BS' scheduling algorithm allows serving packets of the corresponding UE. 

\subsubsection{Absorbing dynamics of channel} In contrast with designs like Vegas~\cite{vegas} which use the overall minimum RTT of a connection during its life time, we use a moving minimum RTT. Monitoring minimum RTT in a moving time window allows us to absorb dynamics of cellular link's capacity, scheduling delays, and in general, different sources of delay in network, without need for having knowledge about the exact sources of those delays, which in practice, are hard to be known at end-hosts. 

\subsubsection{Cellular link as the bottleneck} Based on high demand of cellular-phone users to access different type of contents, new trends and architectures such as MEC~\cite{mec}, MCDN (e.g.~\cite{maxcdn}), etc. have been proposed and used recently to push the content close to the end-users. So, cellular access link known as the \textit{last-mile} becomes the bottleneck even more than before. This insight helps C2TCP's design to concentrate on the delay performance of the last-mile and boost it.

\subsubsection{Isolation of per user queues in cellular networks} Since C2TCP targets cellular networks, it benefits from their characteristics. One of the important characteristics of cellular networks is that usually different UEs get their own isolated deep queues at BS and there is rare competition for accessing queue of one UE by flows of other UEs~\cite{verus,sprout,bufferbloat2}. Mentioned architecture puts BS' scheduler in charge of fairness among UEs using different algorithms such as weighted round robin, or proportional fairness. This fact helps C2TCP to focus more on the delay performance and leave the problem of maintaining fairness among UEs on the last-mile to the scheduler. In addition, it is a reasonable assumption that for cellular end-users there is usually one critical flow using UE's isolated queue at BS when users are running delay sensitive applications such as virtual reality, real time gaming, real time streaming, real time video conferencing, etc. on their smartphones. C2TCP benefits from this fact too. \footnote{If not, users can simply prioritize their flows locally, and send/request the highest priority one first.}

\subsubsection{What if C2TCP shares a queue with other flows} Although the main delay bottleneck in cellular network is the last-mile, there still might be concern about the congestion before the access link (for instance, in the carrier's network). The good news is that in contrast with large queues used at BS, normal switches and routers use small queues~\cite{buffer-size}. So, using well-known AIMD property ensures that the C2TCP will achieve fairness across connections~\cite{aimd} before the flow reaches its isolated deep buffer at BS. In section~\ref{sec_fair}, we show good fairness property of C2TCP in the presence of other flows in such condition.

\subsubsection{Letting loss-based TCP do the calculations} Another helpful insight behind C2TCP is that in contrast with delay-based TCPs, C2TCP does not directly involve packets' delay to calculate the congestion window, but let loss-based TCP, which is basically designed to achieve high throughput~\cite{cubic,reno,newreno,tahoa}, do most of the job. So, instead of reacting directly to every large RTT, definition of ``bad-condition'' helps C2TCP detect persistent delay problems in a time window and react only to them. Therefore, events impacting only a few packets (such as stochastic losses unrelated to congestion) won't impact the algorithm that much. Good resiliency of C2TCP to stochastic packet losses is investigated in section~\ref{sec_loss}.

\subsection{Does C2TCP work in other networks?}
Our design rests on underlying architectures of cellular networks including presence of deep per user buffers at BS, exploiting an scheduler at BS which brings fairness among various UEs at the bottleneck link (last-mile), and low end-to-end control feedback delay (thanks to current technologies and trends such as MEC, MCDN, M-CORD~\cite{mcord}, etc.). Therefore, lack of these structures will impact C2TCP's performance. For instance, for networks with very large intrinsic RTTs, end-hosts absorb the network's condition with a large delay due to the large feedback delay. So, because of that large feedback delay, C2TCP (and any other end-to-end approaches) couldn't catch fast link fluctuations and respond to them very fast. 

%
%

\section{Evaluation}
\label{eval}
In this section, we evaluate performance of C2TCP using extensive trace-driven emulation and compare its performance with existing protocols under a reproducible network condition (source code is available to the community at: \url{https://github.com/soheil-ab/c2tcp}). As our trace-driven emulator, we use Mahimahi \cite{mahi} and use use Iperf application to generate traffic.
	
\textbf{Cellular Traces:} Evaluations are conducted using data collected in prior work (\cite{mahi} and \cite{sprout}) from 5 different commercial cellular networks in Boston (T-Mobile's LTE and 3G UMTS, AT\&T's LTE, and Verizon's LTE and 3G 1xEV-DO) in both downlink and uplink directions.


\textbf{Schemes Compared:} We have implemented C2TCP in Linux Kernel 4.13, on top of Cubic as the loss-based TCP, though any other loss-based TCP variants can be simply used as the base. We use this implementation to compare C2TCP with the state-of-the-art end-to-end schemes including BBR \cite{bbr}, Verus~\cite{verus}, Sprout~\cite{sprout}, and different TCP flavors including Cubic~\cite{cubic}, Vegas~\cite{vegas}, and NewReno~\cite{newreno}\footnote{We saw a bug in LEDBAT's implementation~\cite{ledbat}  which has been confirmed in our conversations with its authors, so we didn't include the result of its performance here}. We also compare C2TCP with CoDel~\cite{codel} an in-network solution which requires to be implemented in the carrier's cellular equipment for downlink queue and in baseband modem or radio interface driver of phones for uplink queue. To do that, we use Cubic at server/client sides and use CoDel scheme as queue management scheme in the network. For C2TCP, unless it is mentioned, we set both \textit{Target} and \textit{Interval} to 100ms. We examine sensitivity of C2TCP to these two parameters in section~\ref{sec_target}.

\textbf{Metrics:} We use 3 main performance metrics for evaluations: average throughput (in short, throughput), average per packet delay (in short, delay), and 95th percentile per packet delay (in short, 95th percentile delay). Average throughput is the total number of bits received at the receiver divided by the experiment's duration. Per packet delay is end-to-end delay which is experienced by a packet from the time being sent to the time being received excluding the propagation delay. Moreover, we investigate the fairness of different schemes. Fairness criterion shows the behaviour of different schemes when there is(are) another normal TCP flow(s) in network. In addition to these metrics, we compare resiliency of different schemes to stochastic packet losses (unrelated to congestion) which might occur in cellular networks.
\begin{figure}[!t]
\centering
\includegraphics[width=0.5\textwidth,height=0.8in]{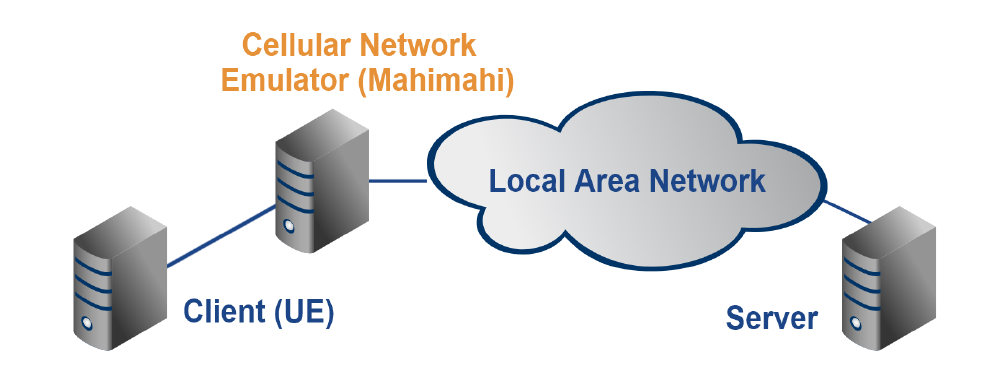}
\caption{Topology used for evaluations}\label{fig_topo}
\end{figure}

\textbf{Topology:} We mainly use 3 entities (equipped with Linux OS) shown in Fig.~\ref{fig_topo} for these evaluations. The first one represents the server, the 2nd one emulates the cellular access channel (and BS) using Mahimahi toolkit, and the 3rd one represents the UE. The RTT is around 40ms. 

\subsection{Comparison with End-to-End Schemes}
\begin{figure*}
\centering
\includegraphics[width=\textwidth,height=\textheight]{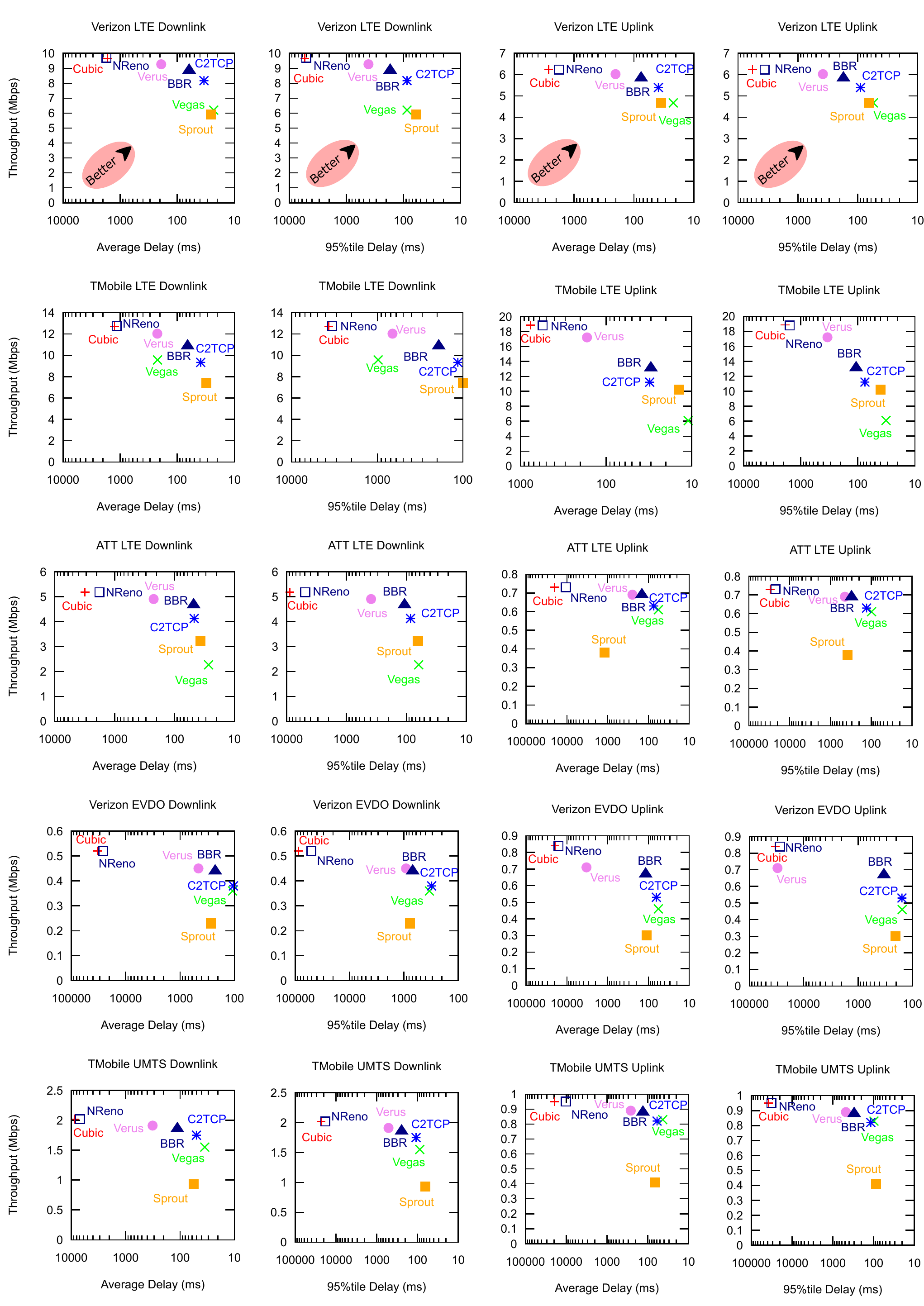}
\caption{Throughput, average delay, and 95th percentile delay of each scheme over traced cellular links (x axis is in logorithmic scale)}\label{fig_overall}
\end{figure*}

Fig. \ref{fig_overall} shows the performance of various end-to-end schemes in our extensive trace-driven evaluations for 5 different measured networks. For each network, there are 2 graphs representing 2 data transfer directions (uplink and downlink), and for each direction there are 2 charts, one showing the average delay and throughput, and the other one illustrating 95th percentile delay and throughput. Schemes achieving higher throughput and lower delay (up and to right region of graphs) are more desirable. 

The overall results averaged across all evaluations have been shown in Table~\ref{table_overall}. C2TCP achieves the lowest average delay and the lowest 95th percentile delay among all schemes, while compromising throughput slightly. For instance, on average, compared to Cubic, C2TCP decreases the average delay more than $200\times$, while compared to Cubic which achieves the highest throughput, it only compromises throughput $0.27\times$.

\begin{table}[!h] \renewcommand{\arraystretch}{1} \caption{Overall results averaged across all traced networks} 
\label{table_overall} \centering
\begin{tabular}{c|ccc}
\hline
& Thr.(Mbps) & Avg. Delay(ms) & 95th\%tile Delay(ms) \\
\hline
C2TCP& 4.235 & \cellcolor{LightBlue}54.1 & \cellcolor{LightBlue}127.2 \\
NewReno& 5.768 & 7688.3 & 16934.5 \\
Vegas& 3.259 & 60.4 & 199.1 \\
Cubic& \cellcolor{LightRed}5.772 & 11015 & 23630.2 \\
Sprout& 3.369 & 185 & 183.4 \\
Verus& 5.408 & 560.7 & 3481.4 \\
BBR& 4.796 & 101.3 & 262.4 \\
\hline
\end{tabular}
\end{table}
Generally, results for different traces in Fig.~\ref{fig_overall} show a common pattern. As expected, Cubic and NewReno achieve the highest throughput among different schemes. The reason is that since they are not sensitive to delay, they simply buildup queues. Therefore, they will achieve higher utilization of the cellular access link when channel experiences good quality. Vegas and Sprout can achieve low delays but they compromise the throughput. Verus performs better than schemes such as Cubic and NewReno and achieves lower average and 95th percentile delays. However, its delay performance is far from the delay performance of Sprout, BBR, Vegas, and C2TCP for almost all traces.  Design of BBR is based on first getting good throughput and then reaching good delays~\cite{bbr}. This explains why BBR can get good throughput while its delay performance is not good. The main idea behind Sprout is to predict the future cellular link's capacity and send packets to the network cautiously to achieve low 95th percentile delay for packets. We observed that Sprout can achieve good delay performance, but it sacrifices throughput. C2TCP tries to achieve low per packet delay while having high throughput. Results confirm that C2TCP achieves the low delay across each of 10 links while maintaining a good throughput performance.\footnote{It is worth mentioning that all experiments have been repeated several times to make sure that the results presented here are not impacted by the random variations.}
\subsection{Comparison with an In-Network Scheme}
\label{eval-codel}
Now, we compare performance of our end-to-end solution C2TCP with CoDel, an in-network solution which is one of the schemes that inspired us. To do that, we add CoDel AQM algorithm to both uplink and downlink queues in Mahimahi and use Cubic at the end hosts. 
Tabe \ref{table_codel} shows the overall results averaged across all traced networks. Using CoDel improves the delay performance of Cubic while degrading its throughput. 
It is worth mentioning that to have in-network solutions such as CoDel, cellular carriers should install these in-network schemes inside their base stations and in base band modem or radio-interface drivers on cellular phones, while an end-to-end scheme like C2TCP only requires updated software at cellular phones, and thus is much easier to be deployed. We show in section~\ref{sec_target} that by changing \textit{target} parameter of C2TCP we can get delay performances better than CoDel's delay performances at the cost of trading throughput.
\begin{table}[h] \renewcommand{\arraystretch}{1} \caption{Overall results averaged across all traced networks} 
\label{table_codel} \centering
\begin{tabular}{c|ccc}
\hline
&Thr.(Mbps) & Avg. Delay(ms) & 95th\%tile Delay(ms) \\
\hline
C2TCP& \cellcolor{LightBlue}4.235 & 54.1 & 127.2 \\
CoDel+Cubic& 4.001 & \cellcolor{LightRed}39 & \cellcolor{LightRed}94.8 \\
\hline
\end{tabular}
\end{table}
\begin{figure}[!t]
\centering
\includegraphics[width=0.48\textwidth,height=3.3in]{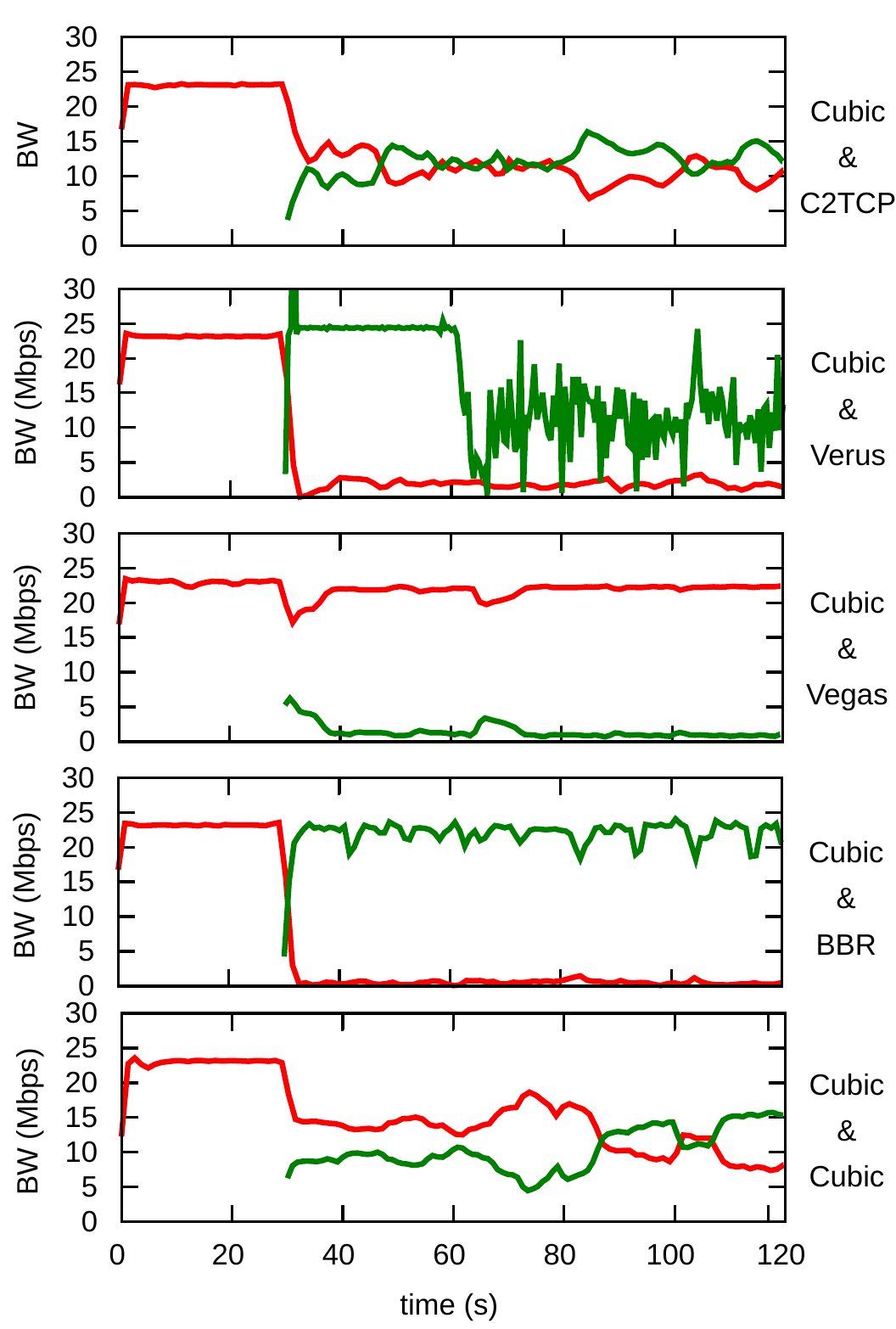}
\caption{Share of bandwidth among Cubic, started at time $0$, and other schemes, started at $30s$ (schemes tested are named on right side of each graph)}\label{fig_fair}
\end{figure}
\subsection{Fairness}
Here, we examine the fairness property of C2TCP. Here, fairness property means that in the presence of other TCP flows, how much fair the bandwidth will be shared among the competing flows. Usually, a scheme that is too aggressive is not a good candidate since it may starve flows of other TCP variants.
\label{sec_fair}
To evaluate the fairness, we send one Cubic flow from one server to an UE. Choosing Cubic as the reference TCP rests on the fact that Cubic is the default TCP in Linux and Android OS which takes more than 60\% of smart phone/tablet market share\cite{android-market}. Then, after 30 seconds, we start sending another flow using the scheme under investigation from another server to the same UE and will report the average throughput gained by both flows through time. When there are unlimited queues in BS, there will be no scheme which can get a fair portion of bandwidth when the queue is already being filled by another aggressive flow~\cite{sprout}. So, to have a fair comparison, as a rule of thumb, we set queue size to the BDP (bandwidth delay product) of the network. Here, the access link's bandwidth and RTT are 24Mbps and 40ms respectively.\footnote{Sprout's~\cite{sprout} main design idea is to forecast the cellular access link capacity using a varying Poisson process, so this scheme won't work properly when link bandwidth is constant. Therefore, to have fair comparison, we don't include performance results of this scheme here.} 

Fig.~\ref{fig_fair} shows the results for different schemes. The results indicate that BBR and Verus are so aggressive and will get nearly all the bandwidth from Cubic flow, while Vegas' share of link's bandwidth cannot grow in the presence of Cubic. 

The main idea of BBR is to set congestion window to the BDP (bandwidth delay product) of the network. To do that, it measures min RTT and delivery rate of the packets. When queue size is at the order of BDP, BBR fully utilizes the queue and will not reserve room for the other flows. Therefore, in our case, cubic flow experiences extensive packet drops and won't achieve its fair share of the bandwidth. Vegas changes its congestion window based on the minimum and the current RTT of the packets. Presence of cubic flow's packets in the queue impact both minimum RTT and current RTT measurements of Vegas. That's why Vegas cannot increase its throughput and get its share of the bandwidth from cubic flow.

\begin{figure}[!t]
\centering
	\begin{minipage}[b]{0.48\linewidth}
		\includegraphics[width=\textwidth,height=0.8in]{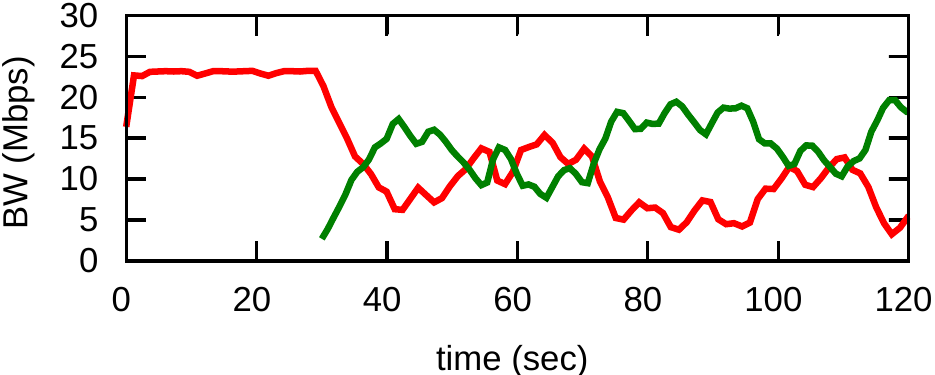}
		\caption{Share of bandwidth among NewReno, started at $0$, and C2TCP}		 
		\label{fig_reno-c2tcp}
	\end{minipage}
	\hfill
	\begin{minipage}[b]{0.48\linewidth}
		\includegraphics[width=\textwidth,height=0.8in]{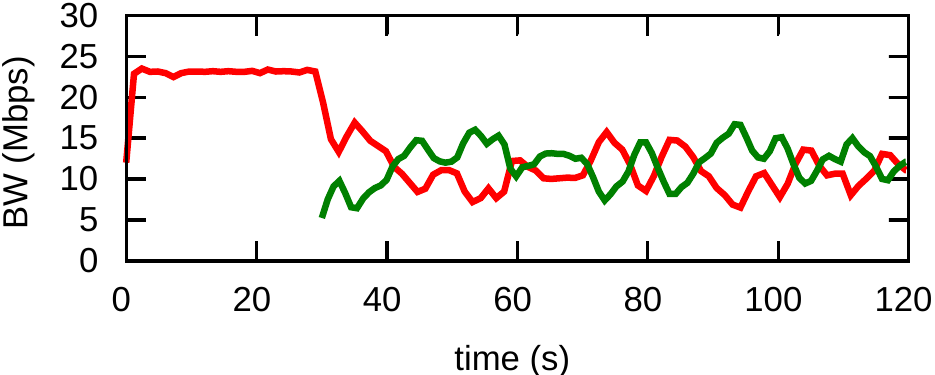}
		\caption{Share of bandwidth among two C2TCP flows}\label{fig_c2-c2}
	\end{minipage}
\end{figure}

In both cases, either being very aggressive or having no aggressiveness, the fairness characteristic of these schemes is not desirable. However, as Fig.~\ref{fig_fair} illustrates, C2TCP  can share the bandwidth with Cubic flow fairly. In our evaluations, C2TCP is implemented over Cubic. To show that C2TCP's fairness property is not because the competing flow in test is Cubic, we replace Cubic flow with a NewReno flow and do the test again. Fig.~\ref{fig_reno-c2tcp} shows the result indicating the same fairness property of C2TCP.

Also, to examine the fairness criterion of the C2TCP in the presence of another C2TCP flow, we use the previous setup and replace the Cubic flow with a C2TCP flow. Results shown in Fig.~\ref{fig_c2-c2} declare that C2TCP is fair to other C2TCP flow in the network. That being mentioned, C2TCP is friendly to other TCP flows and can achieve good fairness property. 
\subsection{Loss Resiliency}
\label{sec_loss}
\begin{figure}[!t]
\centering
\includegraphics[width=0.3\textwidth]{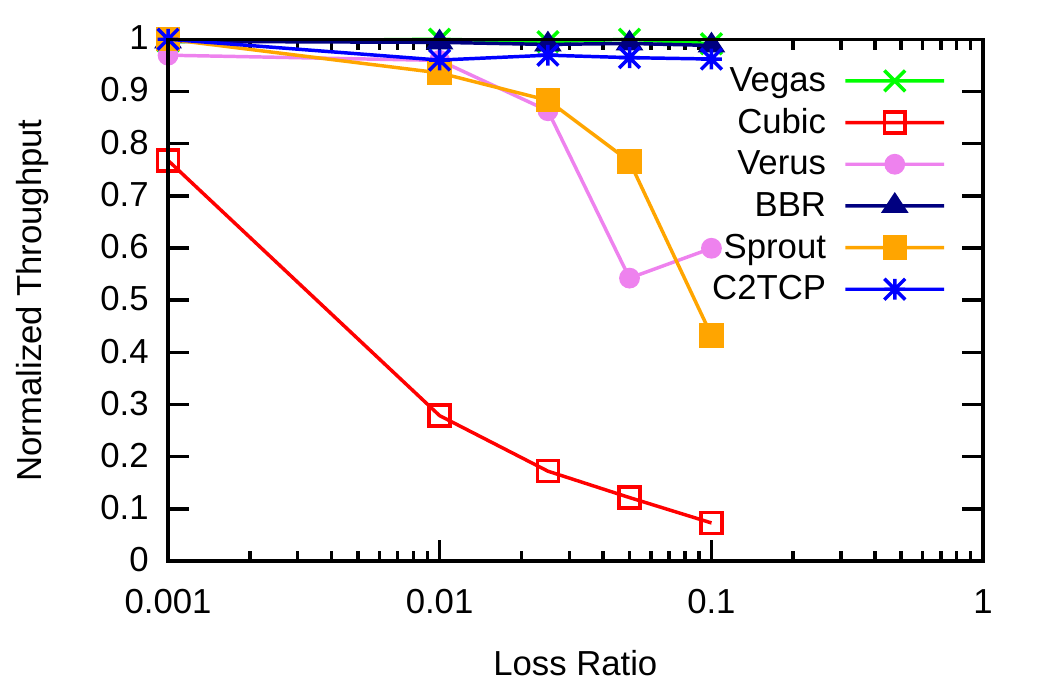}
\caption{Resiliency of schemes to packet losses not caused by congestion}\label{fig_loss}
\end{figure}

Although in cellular networks, different techniques such as HARQ \cite{lte-book} have been used to reduce the impact of stocastic packet losses in access link (which are not caused by congestion), there still could be stocastic packet losses in practice. So, in this section, we investigate the resiliency of different schemes to packet loss not casued by congestion. To that end, we use one of the data traces (Downlink direction of AT\&T's LTE) and simulate Bernoulli packet losses with varying packet loss probabilities. Then, we normalize average throughput of each scheme to the average throughput it sees when there is no loss. This provides us good criterion to see how sensitive each scheme is to the packet losses that are not caused by congestion. Fig.~\ref{fig_loss} shows the results. 

Cubic, a loss-based transport scheme, is sensitive to packet losses and considers them as congestion signals. So, when there are packet losses not due to the congestion, it suffers unwanted slowdowns. However, in parallel with normal mechanism of a loss-based scheme, C2TCP considers delay of packets as the signal of congestion too. When there are packet losses but there is no congestion (which indicates low packet delays i.e. good-condition) C2TCP can speedup the increment process of congestion window using equation~\ref{eq_cwnd} and rectify the unwanted slowdowns. Sprout and Verus both experience decrease in performance specially in high packet losses. However, similar to C2TCP, Vegas and BBR show very good resiliency to packet losses. This is because they both use minimum delay of packets as an extra input for calculating the sending rate, though by using different mechanisms.
\subsection{Impact of Target and Interval}
In this section, we investigate the impact of the only two parameters of C2TCP namely \textit{target} and \textit{Interval} on the performance of C2TCP. We use data trace of downlink direction of AT\&T's LTE network for the evaluations of this section. 
\label{sec_target}
\subsubsection{Target}
\begin{figure}[!t]
\centering
\includegraphics[width=0.45\textwidth,height=1.1in]{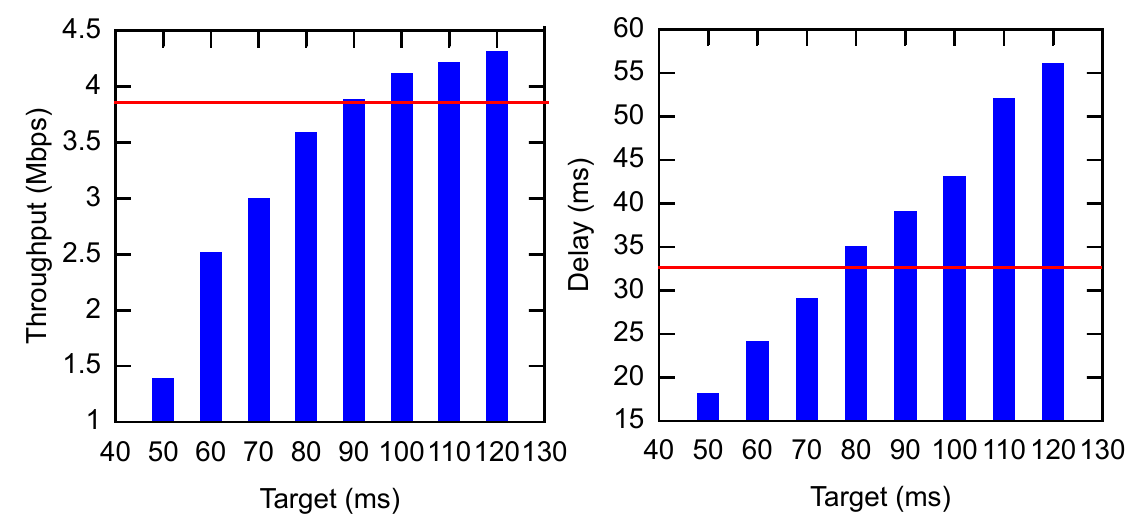}
\caption{Impact of Target Values on Throughput (Left) and Delay (Right). (Red lines shows performance of Codel+Cubic scheme)}\label{fig_target}
\end{figure}

Here, we set the \textit{interval} to $100ms$ and change the \textit{target} from $50ms$ to $100ms$. The average delay and throughput achieved for each setting has been shown in Fig.~\ref{fig_target}. As expected, by changing \textit{target}, an application can achieve a very good balance between throughput and delay. Based on the requirements of different applications, they can change \text{target} via a socket option field. For a balanced performance, we recommend using a target value between $2\times$ to $3\times$ of RTT (here, $RTT=40ms$). This provides enormous flexibility to applications when it is compared to in-network schemes such as RED and CoDel in which a set of queue parameters are set for all applications. Red lines in Fig.~\ref{fig_target} show throughput and delay performance of CoDel when it is used in combination with Cubic for the same scenario. As Fig.~\ref{fig_target} illustrates, C2TCP can be tuned to outperform the performance of CoDel, an in-network approach. 
\subsubsection{Interval}
\begin{figure}[!t]
\centering
\includegraphics[width=0.45\textwidth,height=1.1in]{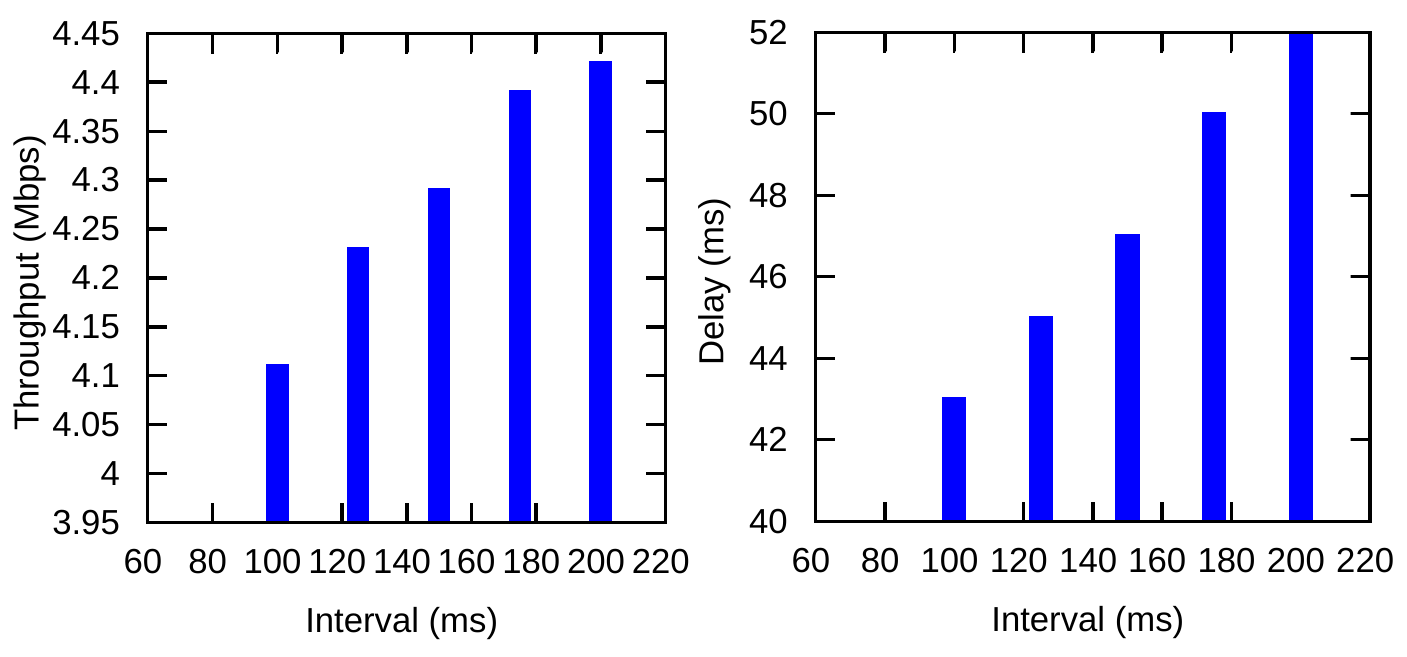}
\caption{Impact of Interval Values on Throughput (Left) and Delay (Right)}\label{fig_interval}
\end{figure}

Now, we set the \textit{target} to $100ms$ and change the \textit{interval} from $75ms$ to $200ms$, and report the average delay and throughput for each setting in Fig.~\ref{fig_interval}. As expected, increasing \textit{interval} increases throughput at cost of delay and vice versa. Generally, we find out that setting \textit{interval} to a few times of RTT is sufficient, though applications can change it using socket options, if they need.

\section{Related Work}
\textbf{End-to-end congestion control protocols:}
Congestion control is always one of the hottest topics with huge studies including numeric variants of TCP. TCP Reno\cite{reno}, TCP Taho\cite{tahoa}, and TCP NewReno\cite{newreno} were among early approaches using loss-based structures to control the congestion window. TCP Vegas\cite{vegas} tries to do congestion control directly by using measured RTTs. TCP Cubic\cite{cubic} changes incremental function of the general AIMD-based congestion window structure, and Compound TCP \cite{compound} maintains two separate congestion windows for calculating its sending rate. BBR \cite{bbr} estimates both maximum bottleneck bandwidth and minimum RTT delay of the network and tries to work around this operation point, though \cite{jaf} has proved that no distributed algorithm can converge to that operation point. Also, LEDBAT\cite{ledbat}, BIC\cite{bic}, and TCP Nice\cite{nice} can be mentioned among other variants. However, all these schemes are mainly designed for a wired network, i.e. fixed link capacities in the network. In that sense, they are not suitable for cellular networks where link capacity changes dynamically and stochastic packet losses exist. 

Among the state-of-the-art proposals targeting cellular networks, Sprout~\cite{sprout} and Verus ~\cite{verus} are worth being mentioned. Sprout introduces a stochastic forecast framework for predicting the bandwidth of cellular link, while Verus tries to make a delay profile of the network and then use it to calculate congestion window based on the current network delay. We have compared C2TCP with most of these schemes in section~\ref{eval}.

\textbf{AQM schemes and feedback-based algorithms:}
Active queue management schemes (such as RED~\cite{red}, BLUE \cite{blue}, and AVQ \cite{avq}) use the idea of dropping/marking packets at the bottleneck links so that end-points can react to these drops later and control their sending rates. It is already known that automatically tuning parameters of these schemes in network is very difficult~\cite{codel,sprout}. To solve that issue, CoDel~\cite{codel} proposes using sojourn time of packets in a queue instead of queue length to drop packets and indirectly signal the end-points. However, even this improved AQM scheme still has an important issue inherited from its legacy ones: these schemes all seek a ``one-setting-fits-all'' solution, while different applications might have different throughput or delay constraints. Even one application can have different delay/throughput requirements during different periods of its life time. 

Also, there are different schemes using feedback from network to do a better control over sending window. Among them, various schemes using ECN\cite{ecn} as the main feedback. Most recent example is DCTCP\cite{dctcp} which changes congestion window smoothly using ECN feedback in datacenter networks. However, DCTCP  similar to other TCP variants is mainly designed for stable links but not highly variable cellular links.

AQM and feedback-based schemes have a common problem: they need changes in the network which is not desirable by cellular network providers due to high CAPEX costs. Inspired by AQM designs such as CoDel and RED, C2TCP provides an end-to-end solution for this issue. Our approach doesn’t require any change/modification/feedback to/from net-
work


\section{Discussion}
\subsubsection{Abusing the parameters} Misusing a layer 4 solution and setting its parameters to get more share of the network by users is always a concern. For instance, a user can change the initial congestion window of loss-based schemes such as Cubic in Linux kernel. Similarly, users can abuse the Target/Interval parameters of C2TCP. Although providing mechanisms to prevent these misuses is beyond the scope of this paper, we think that setting minimum and maximum allowed values for C2TCP's parameters can alleviate the issue. In addition, in TCP, sender's congestion window will be capped to the receiver's advertised window (RcvWnd). Therefore, even by setting the Target value to a very large number in C2TCP, congestion window will be capped to RcvWnd at the end. 
\subsubsection{C2TCP flows with different requirements on one user}
When a cellular phone user runs a delay sensitive application (such as real-time gaming, video conferencing, virtual reality content streaming, etc.), flow of that application is the main interested flow (highest priority one) for the user. Therefore, through the paper, we have assumed that it's rare to have more flows competing with that highest priority flow for the same user. However, in case of having multiple flows with different requirements for the same user, we think that any transport control solution (such as Cubic, Vegas, Sprout, C2TCP, etc.) should be accompanied with prioritization techniques at lower layers to get good results in practice (e.g. \cite{hyline,ffq}). For instance, one simple existing solution is using the strict priority tagging for packets of different flows (by setting differentiated services field in the IP header) and later serve flows based on these strict priorities in the network.
\subsubsection{Setting Target in practice} 
In practical scenarios, instead of setting Target value per application, we could set it per class of applications. In other words, we could let applications choose their application types. Then, C2TCP would set the Target using a table including application types and their corresponding Target values made in an offline manner. 

\section{Conclusion}
We have presented C2TCP, a congestion control protocol designed for cellular networks to achieve low delay and high throughput. C2TCP's main design philosophy is that achieving good performance does not necessarily comes from complex rate calculation algorithms or complicated channel modelings. C2TCP attempts to absorb dynamics of unpredictable cellular channels by simply investigating local minimum delay of packets in a moving time window. Doing that, C2TCP stands on top of an existing loss-based TCP and provides it with a sense of delay without using any network state profiling, channel prediction, or complicated rate adjustments mechanisms. We show that C2TCP outperforms well-known TCP variants and existing state-of-the-art schemes which use channel prediction or delay profiling of network.


\bibliographystyle{IEEEtran}

\end{document}